\documentclass[twocolumn,prc,aps,nofootinbib,numbers,sort&compress]{revtex4}

\usepackage[latin9]{inputenc}
\usepackage{color}
\usepackage{amsmath}
\usepackage{amsthm}
\usepackage{amssymb}
\usepackage{graphicx}
\usepackage[dvipsnames]{xcolor}
\usepackage{ulem,xpatch}

\usepackage{booktabs}
\usepackage{makecell}

\makeatletter

\usepackage{epsfig}
\usepackage{amsthm}
\usepackage{tikz}
\usetikzlibrary{quantikz}

\newcommand{\denis}[1]{\textsf{\color{JungleGreen}{\textsuperscript{DL}(#1)}}}

\usepackage{bm}

\usepackage{float}

\DeclareMathOperator{\Tr}{Tr}

\usepackage{color}

\makeatother

\begin{document}

\title{Restoring symmetries in quantum computing using Classical Shadows}

\author{Edgar Andres Ruiz Guzman }
\email{Andres.Ruiz@ibm.com}
\affiliation{Universit\'e Paris-Saclay, CNRS/IN2P3, IJCLab, 91405 Orsay, France}

\author{Denis Lacroix }
\email{denis.lacroix@ijclab.in2p3.fr}
\affiliation{Universit\'e Paris-Saclay, CNRS/IN2P3, IJCLab, 91405 Orsay, France}

\date{\today}

\begin{abstract}
We introduce a method to enforce some symmetries starting from a trial wave-function prepared on quantum computers that might not respect these symmetries. The technique eliminates the necessity for performing the projection on the quantum computer itself. Instead, this task is conducted as a post-processing step on the system's ``Classical Shadow''. Illustrations of the approach are given for the parity, particle number, and spin projectors that are of particular interest in interacting many-body systems. We compare the method with another classical post-processing technique based on direct measurements of the quantum register. We show that the present scheme can be competitive to predict observables on symmetry-restored states once optimization through derandomization is employed. The technique is illustrated through its application to compute the projected energy for the pairing model Hamiltonian.
\end{abstract}


\maketitle 

\section{Introduction}

\begin{figure*}[htbp]
    \centering
    \includegraphics[width=\linewidth]{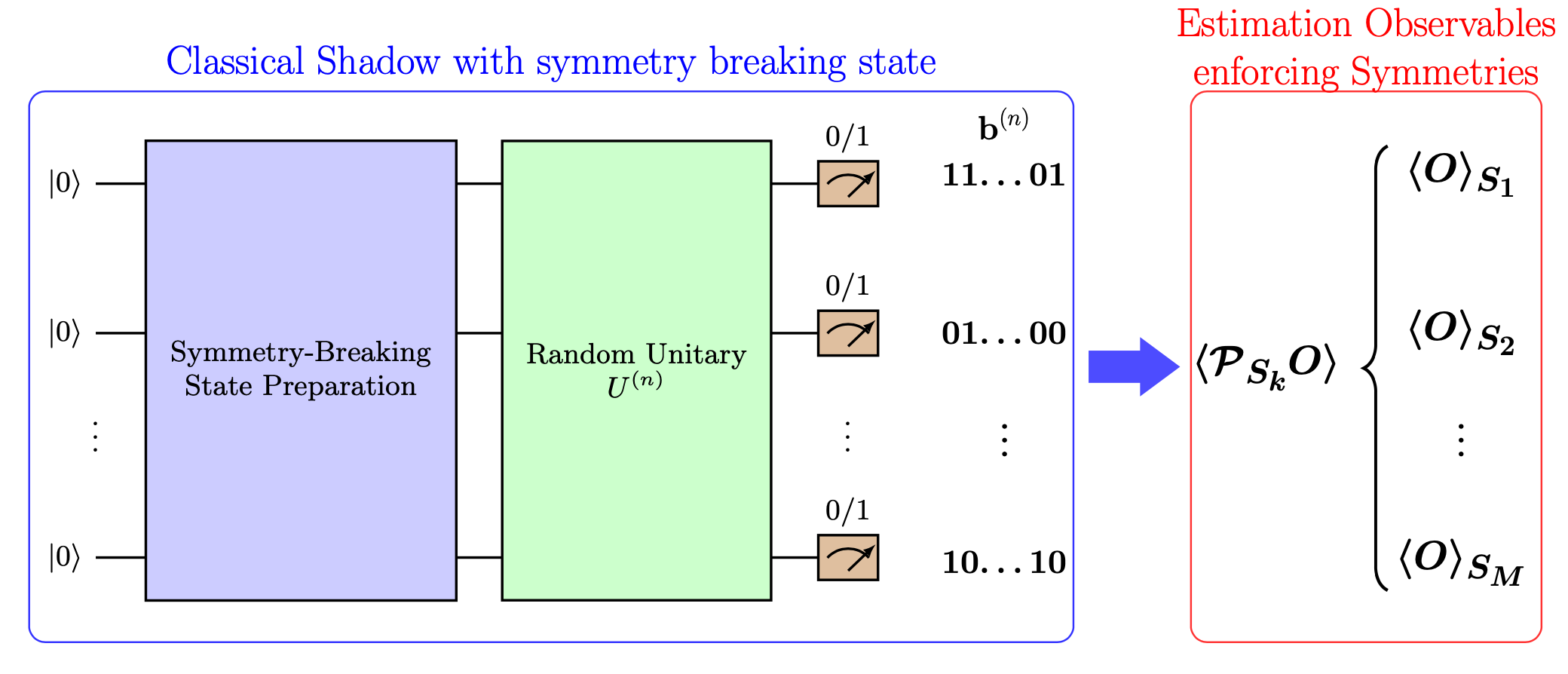}
    \resizebox{\textwidth}{!}{
    \include{fig1}
    }
    \caption{Schematic illustration of the strategy used in the present article to combine classical shadow and symmetry projection. A symmetry-breaking state 
    is prepared on a quantum register. Randomized circuits are implemented, followed by measurements performed to obtain a set of binary numbers that are then used to implement the Classical Shadows technique. In the present work, we use the Classical Shadows method to estimate the expectation value of observables not directly on the (symmetry-breaking) state prepared on the quantum computer but on a different state obtained from it by imposing some symmetries that are not respected by the original state. In the Classical Shadows approach, the symmetry is imposed by explicitly introducing the projector, denoted here by ${\cal P}_{S_k}$. Here, the $\{ S_k\}$ denotes the different eigenvalues of a symmetry operator that can be seen as different channels associated with different sectors of the total quantum Hilbert space. It is worth noting that in this method, an estimation of all the expectation values $\langle O \rangle_{S_k}$ are obtained at the same time once the classical shadow of the symmetry-breaking state has been retrieved. The present quantum-classical strategy based on symmetry-breaking/symmetry restoration, therefore, delegates the delicate task of restoring some symmetries of a problem in classical post-processing.}
    \label{fig:shadow0}
\end{figure*}
Recent advancements in quantum technological platforms have ignited profound interest in the potential for securing real computational benefits from the quantum computing paradigm. While certain quantum algorithms known to offer advantages over their classical equivalents have been established \cite{Nie02,Hid19}, they are currently not implementable on the present Noisy Intermediate-Scale Quantum (NISQ) platforms \cite{Pre18,Bha22}. Considerable progress in quantum computing hardware is required before these methods become viable, possibly anticipating a fault-tolerant era. Nevertheless, substantial research is being invested in exploiting the existing capabilities of NISQ devices in search of the stated advantage. Quantum heuristics \cite{Hog00,Far14}--algorithms with polynomial complexity without guaranteed accuracy--form the central focus of this endeavor, as they enable the customization of requirements to lessen circuit depth and the number of qubits used, leveraging the specifics of the problem at hand. Among these algorithm classes, variational algorithms are distinguished by their adaptability and noise-handling capability. The variational quantum eigensolver (VQE) \cite{Per14,McC16}, owing to its relatively short depth and the potential to tailor an ansatz to a given problem, is one of the most promising contenders within this category.

Enforcing symmetries and associated conservation laws of a physical problem is a decisive aspect to adequately describe 
it on classical or quantum computers \cite{Gro96}. The design of symmetry-preserving quantum circuits might rapidly increase the 
number of quantum gates as the size of the quantum register increases. In the context of many-body systems, circuits enforcing, for instance, particle 
number conservation, total spin symmetry, and/or permutation invariance are illustrated in Refs. \cite{Aru20,Gar20,Sek20,Sek22,Wil11}. Alternatively, as it is well-known in classical computing, for some specific problems, it might be compelling to use wave functions that explicitly break some symmetry of a problem \cite{Nam09,Bee19}. One of the key aspects of using symmetry-breaking ansatz, which is one of the origins of its success in classical computing, is 
that this state can grasp complex correlations/entanglement at much lower numerical costs compared to the cost needed to reach the same level of correlations using symmetry-preserving states. The price to pay is to allow exploring unphysical sectors of the Hilbert space, which might require re-projecting the obtained wave function onto the physical subspace. This step is usually referred to as symmetry restoration  \cite{Rin80,She19}. 

In recent years, with the primary motivation to design accurate ansatz for interacting many-particle systems, an effort has been made to incorporate a symmetry-breaking/symmetry-restoring (SB/SR) technique into the conventional VQE algorithm \cite{Rui22} when symmetries are present (for a review see \cite{Lac23}). Specifically, several quantum algorithms have been proposed to perform the symmetry restoration based, for instance, on the Quantum Phase Estimation \cite{Lac20,Siw21,Rui22}, the concept of quantum Oracle 
and/or the use of Linear Unitary Combination \cite{Rui23}. 

A major obstacle in implementing the symmetry-breaking/symmetry-restoring (SB/SR) methodology today lies in the trade-off between circuit depth and gate operation count. While preparing a symmetry-breaking state can offer a quantitative advantage in terms of circuit depth, this benefit may be partially or entirely offset by the increased number of gate operations required for symmetry restoration, especially when this step is performed on the quantum device itself. This complicates the application of symmetry projections on NISQ devices. A second challenge is that the presence of noise in quantum machines will automatically induce further symmetry-breaking that will only be partially corrected by the projection since the algorithm for projection is also subject to noise.      

Within the NISQ endeavor to minimize the quantum resources necessary for the execution of quantum algorithms, the strategy that seems the most appropriate is to develop potentially highly entangled ansatz based on a symmetry-breaking/symmetry-restoring (SB/SR) technique. This approach allocates to the quantum computer the task of building the symmetry-breaking quantum ansatz, while the symmetry restoration and the estimates of observables with the resulting SR states are relegated to classical post-processing. This would have, in particular, the advantage of taking care of both the symmetry-breaking made on purpose as well as the unwanted one induced by the noise. Let's consider a state \(| \Psi(\bm{ \theta}) \rangle\) built on a digital quantum device, which breaks a specific symmetry \({\cal S}\). Now, let's introduce an operator \(\mathcal{P}_{S}\) that restores the symmetry. This operator achieves restoration by projecting the wave function onto the appropriate subspace of states that adhere to the symmetry \({\cal S}\). If one assumes that this operator can be written as $\mathcal{P}_{S} = \sum_{\alpha} \beta_\alpha V_\alpha$, where the $\{ V_\alpha\}$ are a set of unitary operators. 
Then, the expectation value of any operator $O$ on the symmetry-restored state is given by   
\begin{eqnarray}
\langle O \rangle_{\rm SR} = \frac{\langle\Psi\left({\bm \theta} \right)| \mathcal{P}_{S} O \mathcal{P}_{S}|\Psi\left({\bm \theta} \right)\rangle}{\langle\Psi\left({\bm \theta} \right)|\mathcal{P}_{S}|\Psi\left({\bm \theta}\right)\rangle} = 
\frac{\sum_\alpha \beta_\alpha \langle O V_\alpha \rangle_{\rm SB}}
{\sum_\alpha \beta_\alpha \langle V_\alpha \rangle_{\rm SB}},
\label{eq:obssr}
\end{eqnarray}
where we used the fact that $\mathcal{P}^2_{S} = \mathcal{P}_{S}$, and in the last expression, we implicitly assumed $[O, \mathcal{P}_{S} ]=0$. 
This expression illustrates that the expectation values of the SR states can be written as a combination of several expectation values of the SB that 
can be achieved by a set of measurements performed on circuits involving only the SB state. For many-body systems, the operator $O$ can be directly the 
Hamiltonian $H$. Then the technique can be considered as a practical method to implement SB-SR approach into the conventional Variational Quantum Algorithm (VQE) \cite{Per14,McC16,Kok19,Hug21}. This technique was used, for instance, in Ref. \cite{Kha21} and \cite{Tsu20,Tsu22} to restore the particle number and total spin, respectively. This method was also used in the so-called Post-selected symmetry verification technique for error mitigation \cite{Bon18}. 
Such techniques will be referenced below as {\it implicit} symmetry restoration.

While obtaining access to the projected state could shed light on the nature of the ground state wave function, 
our primary interest during the process lies in extracting the expectation value of a certain number of observables given by Eq. (\ref{eq:obssr}), like the Hamiltonian itself.
This is precisely the objective of recently developed tomography techniques \cite{Nie02,Dar03,Tor23,Aar18,Hua20}, which have demonstrated the capability of retrieving expectation values with significantly diminished resources compared to conventional methods. Additionally, once the variational process is terminated, the study of the projected ground state itself obtained at the minimum of energy 
can be conducted using any of the restoration techniques laid out in \cite{Lac23}.

The purpose of this article is to elaborate a novel method along the same line of implicit symmetry restoration--which involves extracting the projected expectation value of observables without explicitly constructing the projected wave function--grounded in the ``Classical Shadows'' framework \cite{Aar18,Hua20}. 
This process transitions the projection to a fully classical domain, thereby reducing the deployment of quantum resources to the bare minimum necessary for preparing the variational wave function. 

After summarizing selected aspects of the practical implementation of the Classical Shadows approach, we discuss how this approach can be combined with symmetry restoration. The Classical Shadows technique applied to Symmetry restoration is referred to below as the CS-SR method. A schematic illustration of this method based on the original one presented in Ref. \cite{Hua21} is shown in Fig. \ref{fig:shadow0}. 
We will introduce the method and provide examples of its application on parity, particle number, and spin symmetries. Example of uses of the Shadow technique with the VQE in the symmetry-restored subspace are made on the pairing Hamiltonian \cite{Bri05}.
Lastly, we critically analyze the Shadow technique in terms of quantum 
resources/precision with or without derandomization technique \cite{Hua21} and compare it to alternative methods for symmetry 
restoration.

\section{The Classical Shadows technique}

\subsection{General discussion on tomography}

It is possible to extract the expectation of observables over a quantum system defined over $q$ qubits using tools of quantum tomography \cite{Nie02}. A simple approach for these kinds of methods is to reconstruct a system's density matrix $\rho$ by computing expectation values over a tomographically complete set of operators $\mathcal{U}$--a set $\mathcal{U}$ is deemed tomographically complete if, for any pair of distinct density matrices $\rho$ and $\sigma$, there exists an operator $U \in \mathcal{U}$ and a state $|b\rangle$ such that $\langle b |U\sigma U^\dagger |b \rangle \neq \langle b |U\rho U^\dagger |b \rangle$, where $b=\left\{0,1\right\}^{\otimes q}$, and $|b\rangle = |b_{q-1},\dots, b_1,b_0\rangle$. However, this approach has the drawback of an exponential scaling of the number of expectation values with respect to $q$--translated in an exponential number of measurements--needed to fully reconstruct $\rho$, which is subsequently utilized to obtain the observables' expectation values. Additionally, the reconstruction entails an exponential amount of classical memory and computational resources used to store the density matrix and obtain the expectation value of the observables, respectively. Several methods, such as matrix product state (MPS) tomography \cite{Cra10} and neural network tomography \cite{Tor18,Car19} have been proposed to mitigate these constraints under certain conditions. However, for general quantum systems, these methods still require an exponential quantity of samples \cite{Hua20}.

\subsection{Shadow tomography and Classical Shadows}

As a less resource-intensive alternative to these methods, the ``Shadow Tomography'' technique was introduced in \cite{Aar18}. This approach posits that fully reconstructing the density matrix of a quantum system may be unnecessary for specific tasks. Using this method, it is possible to predict properties like expectation values of a set of observables without requiring complete state characterization. Utilizing a polynomial number of state copies, this technique can predict an exponential number of target functions, including fidelity, expectation values, two-point correlators, and entanglement witnesses. The ``Classical Shadows'' method \cite{Hua20} extended this concept to develop an efficient protocol for obtaining a minimal classical sketch $S_{\rho}$ (the classical shadow) of an unknown quantum state $\rho$, which can be used to predict arbitrary linear function values $\left\{o_\alpha\right\}$ associated with the operators $\left\{O_\alpha \right\}$:
\begin{equation}
    o_\alpha = \Tr\left(O_\alpha \rho \right),  \qquad 1\leq \alpha \leq M.
\end{equation}

The creation of the classical shadow of a state built through a circuit is based on the repetition of a procedure based on a set of random unitary operations applied to the circuit prior to measurements. 
This results in a set of measurements/events labeled by $n=1,N_{\rm Mes}$. This, in turn, leads to the following scheme:
\begin{enumerate}
    \item Create the state associated to a density $\rho$ on the quantum register with $q$ qubits 
    \item Apply a unitary transformation $U^{(n)}$ to the quantum state $\rho$, i.e.
     \begin{equation}
        \rho \rightarrow U^{(n)}\rho [U ^{(n)}]^\dagger.    
     \end{equation}
     The $U^{(n)}$ corresponds to a randomly selected unitary operation taken in a properly chosen set of tomographically complete unitary operations 
     that transforms the basis from the computation basis to a new basis where the measurement will be made. Usually, when selecting the random unitaries $U$, we consider two distinct ensembles: either tensor products comprised of random single-qubit Clifford circuits or random $q$-qubit Clifford circuits.
     \item Measure the register, leading to a classical bitstring ${\bf b}^{(n)} = b^{(n)}_{q-1} \cdots b^{(n)}_0$ that is stored on a classical computer. This bistring is associated to a pure state 
     density $| {\bf b}^{(n)}  \rangle \langle {\bf b}^{(n)} |$, written in the randomly rotated register eigenbasis.
     \item In order to write the density in a common basis, each density is transformed back to the original computational basis 
     by performing the inverse transformation, leading to a set of simple pure-state densities:
     \begin{eqnarray}
         r^{(n)} &=& [U ^{(n)}]^\dagger | {\bf b}^{(n)}  \rangle \langle {\bf b}^{(n)} |U^{(n)}.
     \end{eqnarray}
     Considering that the unitaries $U^{(n)}$ can be efficiently implemented on a classical computer over the bitstring, a classical description $r^{(n)}$ is stored in classical memory. 
\end{enumerate}
This procedure can be regarded as a generator of events, leading to a set of densities that depends on the original quantum state $\rho$. As was shown in Ref. \cite{Hua20,Hua21}, and summarized below, 
by performing a classical average over these densities, provided that sufficient events are generated, one can compute the quantum expectation value of observables taken on the original quantum state.  

\subsubsection{Expectation values as a classical average}

\begin{figure}
    \centering
    \includegraphics[width=\columnwidth]{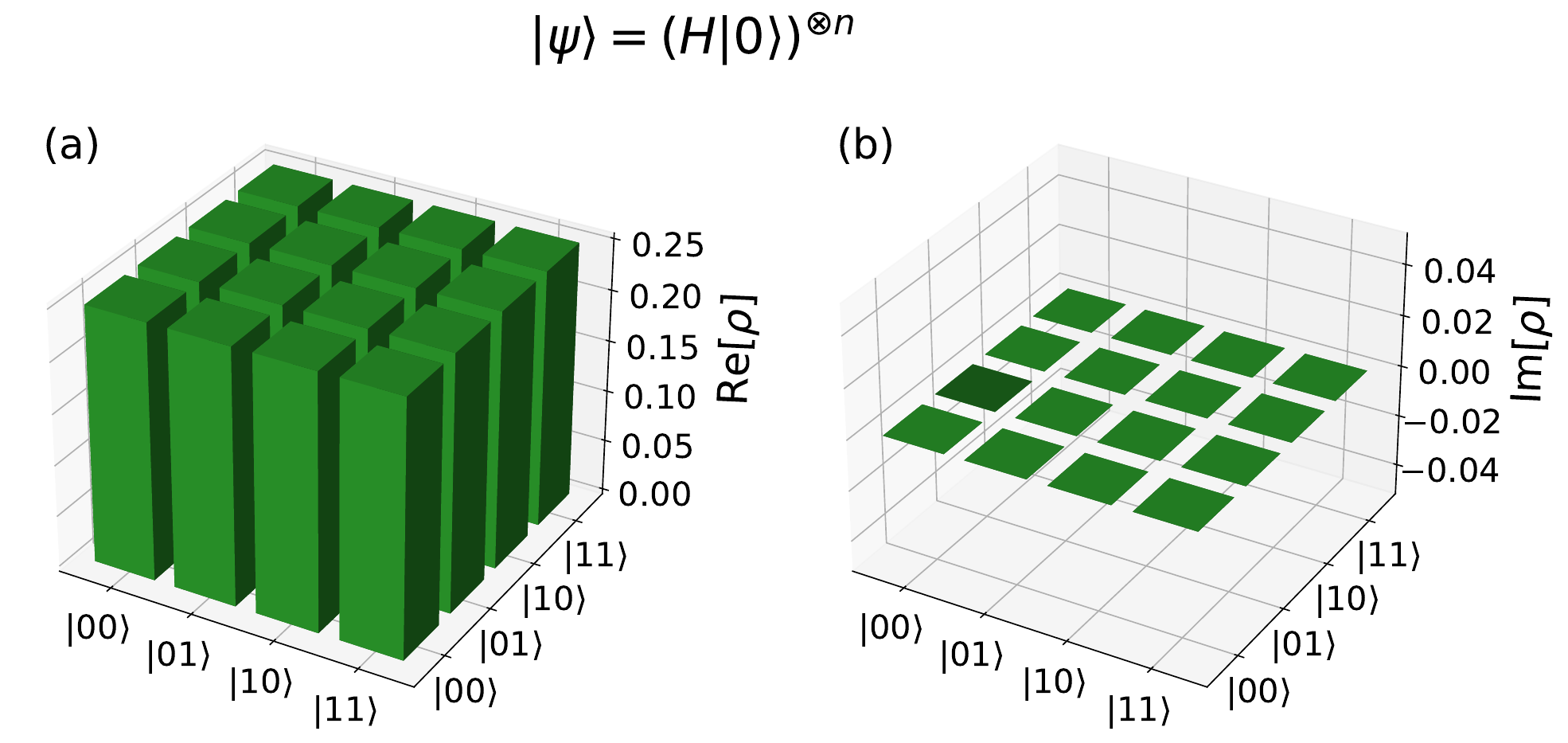}
    \includegraphics[width=\columnwidth]{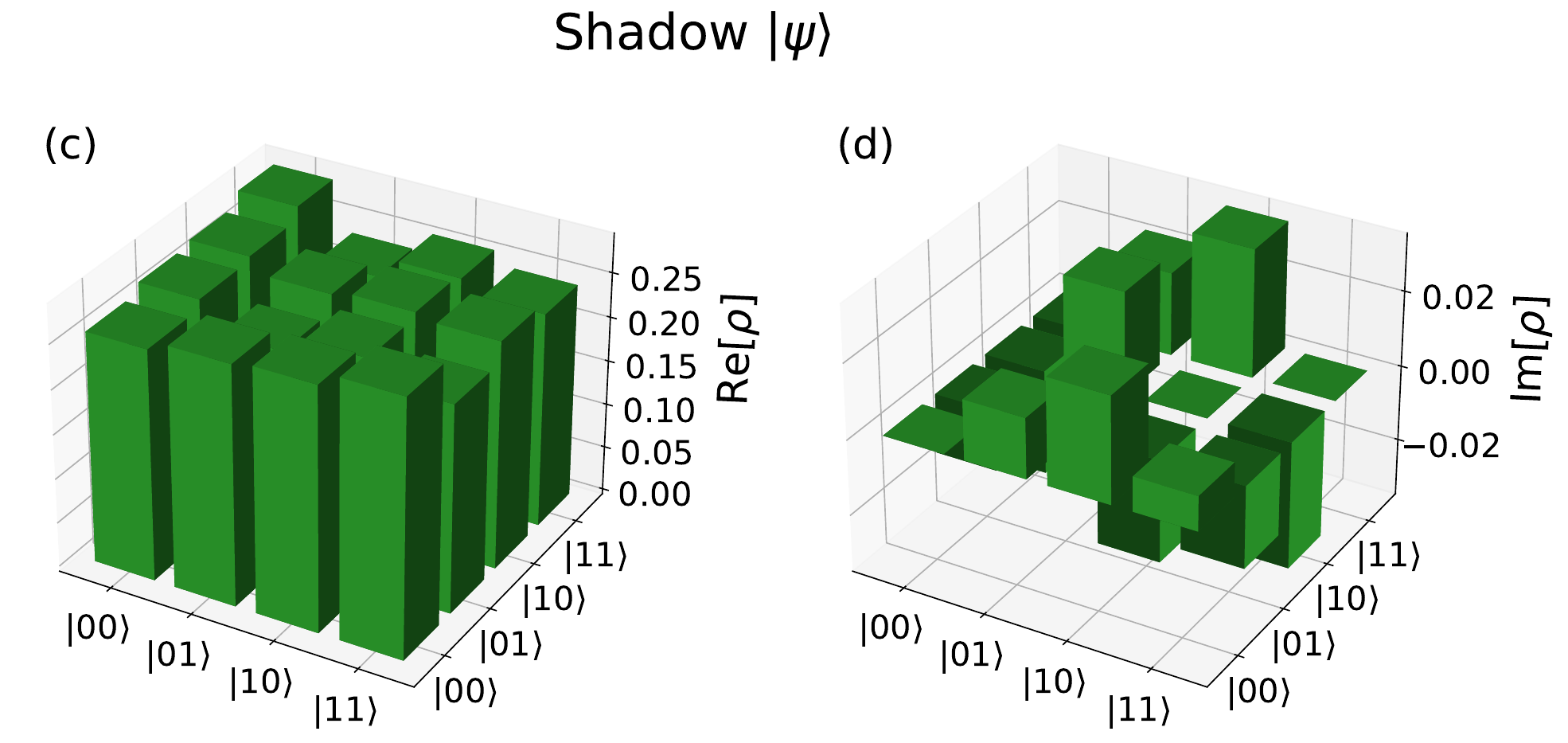}
    \caption{Illustration of the shadow tomography of the two qubits density matrix $\rho \equiv H_1H_0|00\rangle\langle 00|H_0 H_1$, where $H_{j}$ denotes the Hadamard gate applied to the $j$-th qubit. In panels     (a) and (b) are shown the real and imaginary parts of the density $\rho$. Panels (c) and (d) display, respectively, the real and imaginary parts of the reconstructed shadow approximation obtained using 1000 shots. The difference between the exact and approximated density stems from the finite number of shots. Results shown here and hereafter have been obtained using the Qiskit IBM software assuming a fault-tolerant quantum computer \cite{Qis21}. }
    \label{fig:rho_reconstruction}
\end{figure}

Let us denote by $\mathbb{E}$ the average over both the choice of $U^{(n)}$ and the outcomes of the measurement, such that we have
\begin{eqnarray}
    \mathcal{M}\left(\rho\right) &\equiv& \frac{1}{N_{\rm Mes}} \sum_{(n)=1}^{N_{\rm Mes}} r^{(n)} = \mathbb{E}[r^{(n)}]. 
\end{eqnarray}
$\mathcal{M}$  can be viewed as a quantum channel that maps the original $\rho$ to a set of ``classical'' representations 
$r^{(n)}$.
The form of the quantum channel depends on the ensemble of random unitary transformations $\mathcal{U}$. If this ensemble is tomographically complete, we can invert the quantum channel to reconstruct the state:
\begin{equation}
    \rho = \mathbb{E}\left[\mathcal{M}^{-1} \left( r^{(n)} \right)
\right].
    \label{eq:rho_from_shadow}
\end{equation}
Each term in the average can be viewed as a single classical snapshot $\hat{\rho}^{(n)}$:
\begin{equation}
    \hat{\rho}^{(n)} = \mathcal{M}^{-1}\left(r^{(n)}  \right)  = \mathcal{M}^{-1}    
    \left([U ^{(n)}]^\dagger | {\bf b}^{(n)}  \rangle \langle {\bf b}^{(n)} |U^{(n)} 
    \right).
    \label{eq:single_snapshot}
\end{equation}
A collection of $M$ snapshots $\left\{\hat{\rho}^{(1)} , \dots, \hat{\rho}^{(M)} \right\}$ is termed a ``classical shadow'' 
of size $M$.
We show in Fig. \ref{fig:rho_reconstruction} a simple illustration of the use of the classical shadow technique to obtain an approximation for the full density of a two qubits register.

Given that the average of the classical shadow approximates the quantum state $\rho$ (see Eq. \eqref{eq:rho_from_shadow}), it becomes feasible to estimate the expectation value of any observable $O$ using an empirical mean over the individual snapshots:
\begin{equation}
\langle O \rangle = \frac{1}{M}\sum_{n=1}^M\Tr\left(O\hat{\rho}^{(n)} \right).
\label{eq:mean_shadow}
\end{equation}

\subsubsection{Estimation of Pauli observables using Classical Shadows}
\label{sec:energy_shadow}

Suppose we have an observable $O$ expressed as a weighted sum of Pauli chains:
\begin{equation}
    O = \sum_\alpha \gamma_\alpha O_\alpha,
    \label{eq:ham_sum}
\end{equation}
where each $O_\alpha$ is of the form $O_{\alpha} = \bigotimes_{j=0}^{q-1} P^\alpha_j$ and $P^\alpha_j$ identifies either to the identity or one of the 
Pauli matrices, i.e., $P^\alpha_j \subset \left\{I_j,X_j,Y_j,Z_j\right\}$. $\{ \gamma_\alpha\}$ is a set of constant coefficients.
In the following, since the expectation value $\langle O \rangle$ obtained using the sum in Eq. (\ref{eq:ham_sum}) can be easily made classically after computing the expectation values associated with the operators $O_\alpha$, we will simply assume that the operator $O$ identifies with one of the Pauli chain $O_\alpha$ and omit the $\alpha$.

Given the form of the operators $O$, to estimate the expectation values $\langle O \rangle$, it is advantageous to perform the measurements in the basis associated to the 
Pauli operators. 
Thus, we should consider the tomographically complete ensemble associated with the Pauli matrices 
$X$, $Y$, and $Z$ such that:
\begin{equation}
    U^{(n)} = 
    \bigotimes_{j=0}^{q-1} U^{(n)}_j
    \quad {\rm with} \quad U^{(n)}_j \subset \left\{H_i,H_j S^\dagger_j, I_j\right\},
\end{equation}
given that we have the following relations (omitting $j$ for compactness):
\begin{equation}
    X = HZH, \qquad Y = SHZHS^\dagger, \quad {\rm and} \quad Z = IZI.
\end{equation}
When considering this ensemble, 
the inverted quantum channel can be procured from the inverse of every single qubit's quantum channel, i.e., $\mathcal{M}^{-1} = \bigotimes_{j=0}^{q-1}\mathcal{M}^{-1}_j$. Every single snapshot in Eq. \eqref{eq:single_snapshot} then takes the form: 
\begin{eqnarray}
   \hat{\rho}^{(n)} = \bigotimes_{j=0}^{q-1} 
   \left(3 U^{(n)\dagger}_j
   |b^{(n)}_j\rangle \langle b^{(n)}_j|U^{(n)}_j-I_j\right),   
\end{eqnarray}
where, for each qubit, $|b^{(n)}_j\rangle \langle b^{(n)}_j|$ is either $|0_j \rangle \langle 0_j |$ 
or $|1_j \rangle \langle 1_j |$. Introducing the compact notation $r^{(n)}_j = U^{(n)\dagger}_j
   |b^{(n)}_j\rangle \langle b^{(n)}_j|U^{(n)}_j$ such that $r^{(n)} = \bigotimes_{j=0}^{q-1} r^{(n)}_j$,
   the computation of the expectation value of a given Pauli chain 
over every snapshot $\hat{\rho}$ takes the form:
\begin{eqnarray}
    \Tr\left[O\hat{\rho}^{(n)}\right] &=& \prod_{j=0}^{q-1}\Tr\left[P_j\left(3 r^{(n)}_j-I_j\right)\right],\label{eq:expectation_observable}
\end{eqnarray}
due to the fact that the $P_j$ is the identity or a Pauli operator, $U_j$ is the unitary operation we mention before, and $| b^{(n)}_j \rangle \subset 
\{  | 0_j \rangle , | 1_j \rangle \}$. The trace for each qubit can only take particular values reported in the table \ref{tab:values_trace}. This table shows that, as soon as at least one of the $U_j$ does not correspond to its associated operator $P_j$, the entire product cancels out. This distinctive property simplifies the estimation of an observable's expectation value into merely counting the number of ``compatible measurements'' in the classical shadow, compatible meaning that all the $P_j$ that differs from the identity in $O=\bigotimes_{j=0}^{q-1} P_j$ match the measurement protocol. Such a feature is particularly attractive for applications since it can significantly reduce the number of measurements required to estimate an observable. This distinctive property leads to a significant reduction in the number of measurements required to reach a certain accuracy in estimating the expectation values as it allows the formulation of the ``derandomization'' technique, discussed in Ref. \cite{Hua21}.

\begin{table}
    \centering
\[\setcellgapes{3pt}\makegapedcells
\begin{tabular}{c|cccc}
\hline
Operator $U^{(n)}_j$ &  \multicolumn{4}{c}{Operator $P_j$}\\
\cline{2-5}
& $I $ & $X$ & $Y$ & $Z$\\
\hline  
$H$ & $~\left(1,1\right)~$ & $~\left(+3,-3\right)~$ & $(0,0)$ & $(0,0)$\\
$HS^{\dagger}$ & $\left(1,1\right)$ & $(0,0)$ & $~\left(+3,-3\right)~$ & $(0,0)$\\
$I$ & $\left(1,1\right)$ & $(0,0)$ & $(0,0)$ & $~\left(+3,-3\right)~$\\
\hline
\end{tabular}
\]
 \caption{Potential outcomes of the function $\Tr \bigl[ P_j\bigl(3r^{(n)}_j-I_j\bigr)\bigr]$, given a specific unitary operator $U^{(n)}_j$ and operator $P_j$. 
 The results correspond to the measurement states $\left(|0_j\rangle, |1_j\rangle\right)$ of the resultant qubit.}
\label{tab:values_trace}
 \end{table}

\section{Classical shadow with projections}

\subsection{General discussion}
\label{sec:general_discussion_method}

The calculation of projected expectation values associated with a particular Pauli string \(O = \bigotimes_{j=0}^{q-1} P_j\) (which can be part of the decomposition of the Hamiltonian \(H = \sum_{k=0} \alpha_k O_k\)) for the projectors discussed in this paper is conducted by utilizing the property that all of the presented projectors can be written as a linear combination of unitary operators, which themselves can be expressed as a tensorial product of single-qubit operators. To understand how this operates, let's consider a generic projector expressed as:

\begin{equation}
    \mathcal{P} = \sum_k \beta_k R_k,
\end{equation}

with the $R_k = \bigotimes_{j=0}^{q-1} G_k^j$ and $G_k^j$ arbitrary single qubit gates. We know that any single qubit gate $G$ can be expressed as a linear combination of Pauli matrices, i.e.:

\begin{equation}
    G = \sum_m \alpha_m P'_{m},
\end{equation}
with $\alpha_m\in \mathbb{C}$ and $P'_m\in \{I,X,Y,Z\}$. To obtain the projected expectation value, we apply the projector to the Pauli string $O$ and consider its expectation value over every snapshot:

\begin{equation}
    \Tr\left[O\mathcal{P}\hat{\rho}^{\left(n\right)}\right] = \sum_k \beta_k \Tr\left[OR_k\hat{\rho}^{\left(n\right)}\right].
    \label{eq:projected_expectation_value}
\end{equation}

Using the properties of the trace, each trace in the sum can be obtained as:

\begin{equation}
    \Tr\left[OR_k\hat{\rho}^{\left(n\right)}\right] = \prod_{j=0}^{q-1}\left\{\sum_{m=0}^{3}\alpha_m^{jk}\Tr\left[P_jP'_m\left(3r^{\left(n\right)}_j-I_j\right)\right]\right\}.
\end{equation}

The aforementioned equation suggests that the projected expectation values can be attained by utilizing the same traces as computed in Eq. \eqref{eq:expectation_observable}, albeit considering the Pauli matrix product $P_jP'_m$ in place of merely $P_j$. This methodology applies not only to the projector but also to any operator expressible as a linear combination of unitaries, each being tensor products of single-qubit gates. As will be demonstrated in the particle number projector, many of the coefficients $\alpha_m^{jk}$ often either equate to zero or exhibit similarity across qubits. Importantly, all requisite information concerning the specific symmetry eigenvalue intended for projection is encapsulated within the coefficients $\beta_k$. Upon acquiring all values of $\Tr\left[R_kO\hat{\rho}^{\left(n\right)}\right]$ from the quantum computer, access to the projected expectation values across all symmetry eigenspaces is simultaneously granted, as depicted schematically in Fig. \ref{fig:shadow0}.

\subsection{Parity projection by classical shadow}

\subsubsection{Positive and negative parity probabilities }

We consider here a state $| \Psi \rangle$ that is prepared on a quantum register of $q$ qubits and that is not necessarily 
an eigenstate of the parity operator, defined as $\Pi = \bigotimes_{j=0}^{q-1} Z_j$. This operator has two eigenvalues, $+1$ and $-1$, corresponding to even and odd parity states, respectively. Any state can be decomposed as:

\begin{eqnarray}
    | \Psi \rangle &=& | \Psi_+ \rangle + | \Psi_- \rangle \nonumber 
\end{eqnarray}
where $| \Psi_\varepsilon \rangle = {\cal P}_{\varepsilon}| \Psi \rangle$ correspond to the two non-normalized even ($\varepsilon=+$) and odd ($\varepsilon=-$) components
of the wave-function, while ${\cal P}_{\varepsilon= \pm}$ are the two associated projectors. These projectors can be expressed in terms of the parity operator itself \footnote{Note that one can also use the particle number operator: 
\begin{eqnarray}
N = \frac{1}{2} \sum_j (Z_j - I_j) \label{eq:partnumb}
\end{eqnarray}
to express the parity projector. An alternative form of the projector becomes:
\begin{eqnarray}
    {\cal P}_\varepsilon &=&\frac{1}{2} \left[ I + \varepsilon e^{i \pi N} \right]
\end{eqnarray} 
}

\begin{eqnarray}
    {\cal P}_\varepsilon &=& \frac{1}{2} \left[ I + \varepsilon\Pi\right], 
    \label{eq:parity_projector}
\end{eqnarray}
or, 
\begin{eqnarray}
    {\cal P}_\varepsilon &=& \frac{1}{2} \left[ I + e^{i\frac{\pi}{2}\left(\Pi-\varepsilon\right)} \right], 
    \label{eq:parity_projector_2}
\end{eqnarray}
with $\varepsilon=+1$ (resp. $-1$) associated with the even (resp. odd) parity projection. The implementation of the parity operator $\Pi = \bigotimes_{j=0}^{q-1} Z_j$ consists of the straightforward application of $Z$ gates on each qubit. On the other hand, the operator associated with the second form of the projector, $e^{i\frac{\pi}{2}\Pi}$, can be implemented with a linear complexity of $\mathcal{O}(q)$, utilizing up to $2q$ CNOT gates and a single Z-rotation gate. This complexity is understood by recognizing the operator $e^{ia\bigotimes_{j=0}^{q-1}Z}$ as the exponential of the Walsh operator $w_{2^q}=\bigotimes_{j=0}^{q-1}Z$. For a more comprehensive discussion and, in particular, instructions on constructing the operator $e^{ia\bigotimes_{j=0}^{q-1}Z}$, we refer the reader to \cite{Wel14}. Given the simplicity of implementing the parity operator, we will adopt this form over the second one in the rest of the paper.

\begin{figure}
    \centering
    \includegraphics[width=\columnwidth]{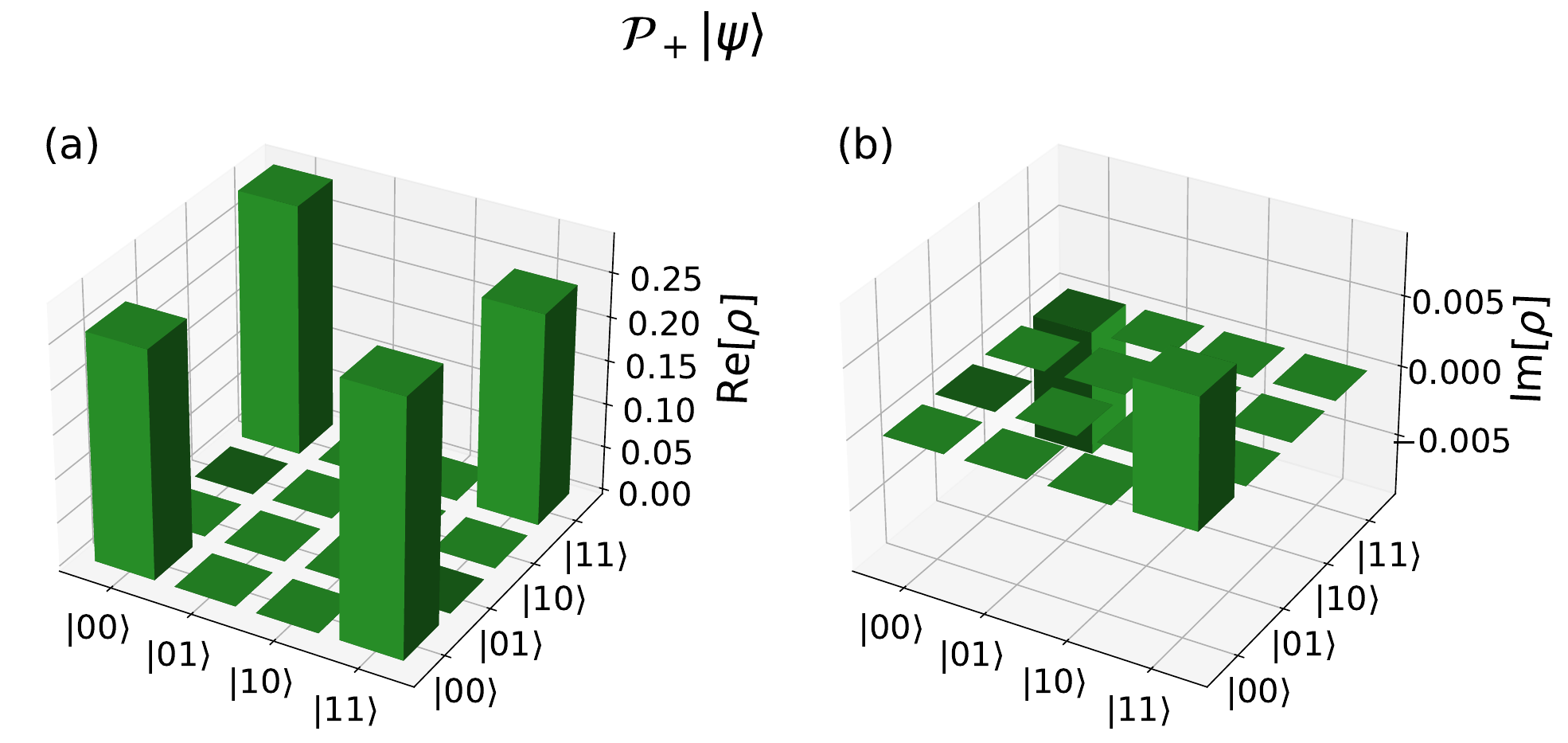}
    \includegraphics[width=\columnwidth]{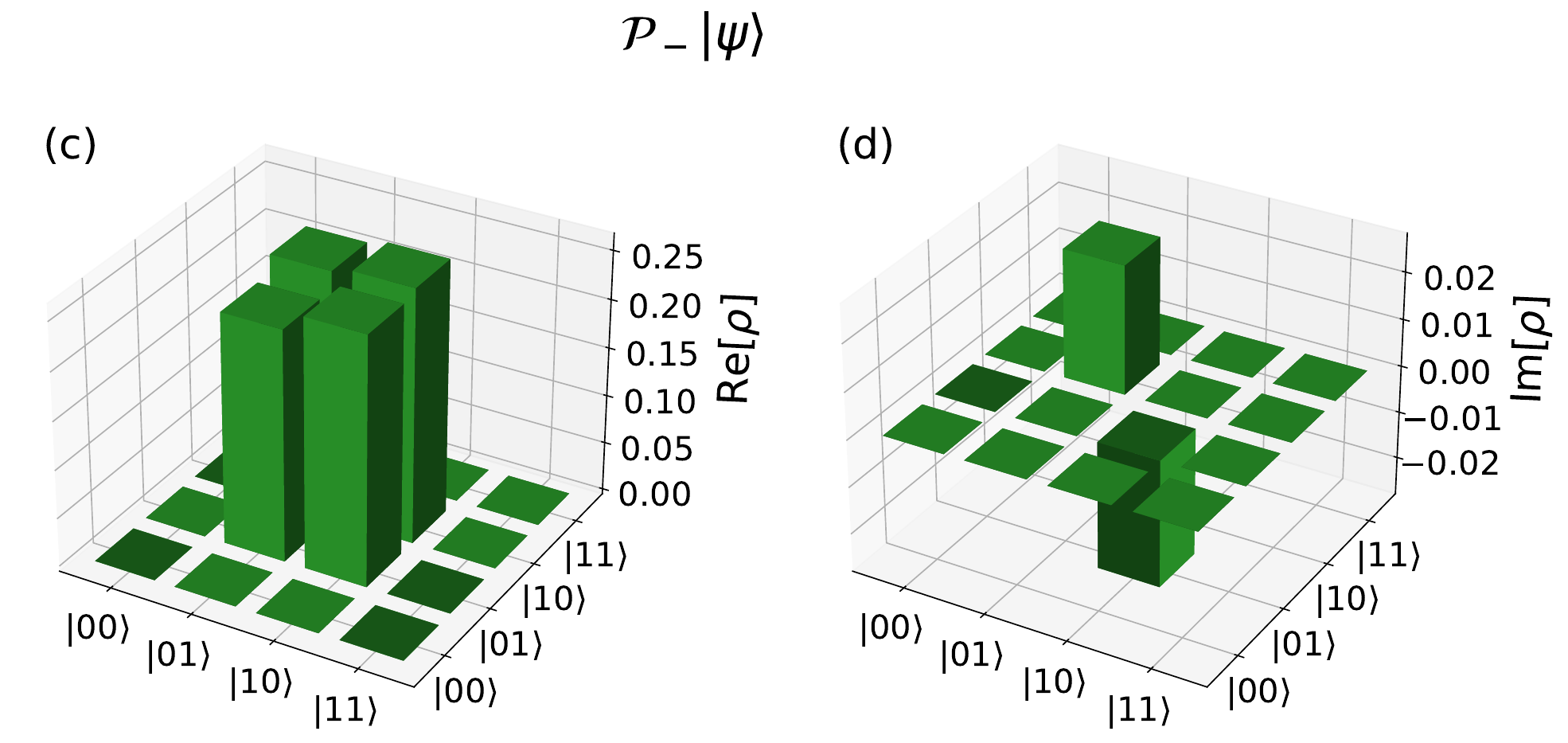}
    \caption{Shadow Tomography of the parity-projected density matrix used in Fig. \ref{fig:rho_reconstruction} using the projector in Eq. (\ref{eq:parity_projector}). Panels (a) and (b) [resp. (c) and (d)] show the real and imaginary parts of the density matrix projected on positive parity (+) [resp. negative parity (-)]. The approximate densities have been obtained using 1000 shots.}
    \label{fig:tomography_projection}
\end{figure}

We show in Fig. \ref{fig:tomography_projection} the approximate densities obtained by combining the Shadow technique and the 
projector on different parities starting from the original mixed-parity density shown in Fig. \ref{fig:rho_reconstruction}. 
In practice, such an image has been obtained by application of the projector in Eq. \eqref{eq:parity_projector} over each snapshot of the form in Eq. \ref{eq:single_snapshot}, i.e., $\mathcal{P}_{\varepsilon}\hat{\rho}^{\left(n\right)}\mathcal{P}_{\varepsilon}$.

\begin{figure}
    \includegraphics[width=\columnwidth]{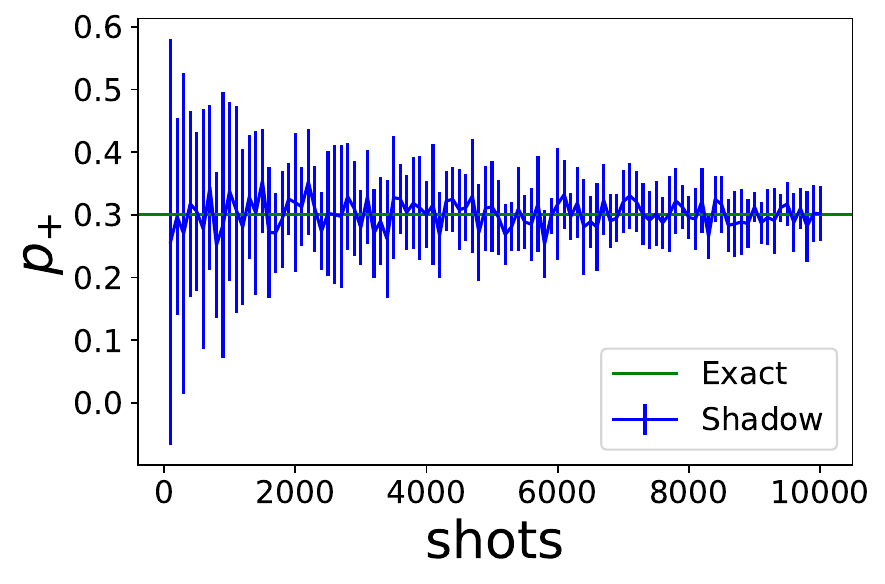}
    \includegraphics[width=\columnwidth]{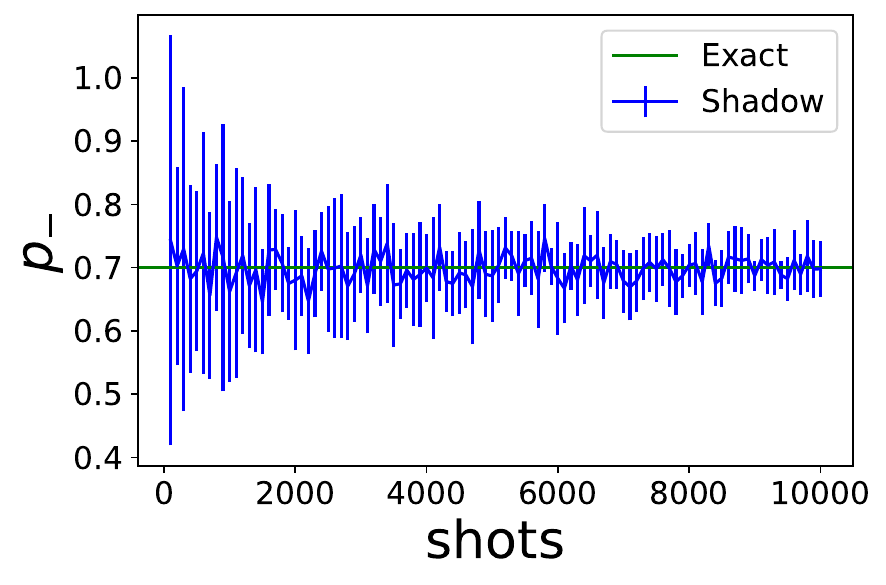}
    \caption{Illustration of the convergence of the probability $p_+$ (top) and $p_-$ (bottom) defined in Eq. (\ref{eq:pparity}) obtained using the Classical Shadows technique combined with symmetry restoration for the parity case. The green solid line indicates the exact result. Estimated probabilities are shown as a function of the number of measurements (or \textit{shots}) for $q= 4$ qubits and for an initial state with $p_{+}=0.3$ and $p_{-}=0.7$, respectively. Each point corresponds to the mean of 10 runs of the corresponding number of shots. The error bars correspond to the standard deviation of those 10 runs.}
    \label{fig:parityprob_shadow}
\end{figure}

In order to illustrate the predictive power and the convergence of the shadow technique with the number of shots, we show  
in Fig. \ref{fig:parityprob_shadow} the odd and even probabilities defined by:
\begin{eqnarray}
    p_{\varepsilon} &=& {\rm Tr} \left( {\cal P}_\varepsilon \rho \right), \label{eq:pparity}
\end{eqnarray}
with $\varepsilon = +$ (resp $-$) for even (resp. odd) parities. 
We observe a perfect convergence to the exact probabilities. Note that 
no specific optimization of the Classical Shadows tecnique has been made to obtain this result. A dedicated discussion is made in section \ref{sec:optimization} on the possible way to improve the precision while reducing the number of shots.   

\subsubsection{Expectation values of observables after projection}

As an illustration of using the classical shadows for observables, we consider the pairing Hamiltonian \cite{Bri05} that was extensively used recently to benchmark symmetry restoration techniques on quantum computers \cite{Rui22,Rui23}. This Hamiltonian 
can be written in second quantization as:
\begin{eqnarray}
H & = &\sum_i\varepsilon_i (a^\dagger_i a_i + a^\dagger_{\bar i} a_{\bar i} ) 
- g \sum_{ij} \hat P^\dagger_i \hat P_j,   \label{eq:hampairing}
\end{eqnarray}
where $(a^\dagger_i,a^\dagger_{\bar i})_{i=0,q-1}$ corresponds to a set of creation/annihilation operators associated with pairs of time-reversed single-particle states labeled by $i$ and $\bar i$. Here $P^\dagger_{i} = a^\dagger_i a^\dagger_{\bar i}$ are the so-called pair creation operators.  
This Hamiltonian is an archetype of problems where it is advantageous to break a symmetry (here the $U(1)$ symmetry associated with particle number conservation, to treat the non-perturbative correlations between particles), especially at large values of the coupling strength $g$. We consider below the case of equidistant doubly degenerated single-particle levels with $\varepsilon_i =i \Delta \varepsilon$ with $i=0, q-1$. When even particle number is considered 
with no broken pairs, such Hamiltonian can be directly encoded on $q$ qubits using the Jordan-Wigner transformation (JWT) with the pair creation operator \cite{Kha21,Rui22}. To estimate the parity-projected energy defined as:
\begin{eqnarray}
    E_{\pm} &=& \frac{1}{p_{\pm}}{\rm Tr}({\cal P}_{\pm} H \rho),
\end{eqnarray}
we use the procedure layout in section \ref{sec:general_discussion_method}. In the equation, $p_{\pm}$ corresponds to the probabilities defined in Eq. \eqref{eq:pparity} and illustrated in Fig. \ref{fig:parityprob_shadow}.

\begin{figure*}
    \centering
    \includegraphics[width=\textwidth]{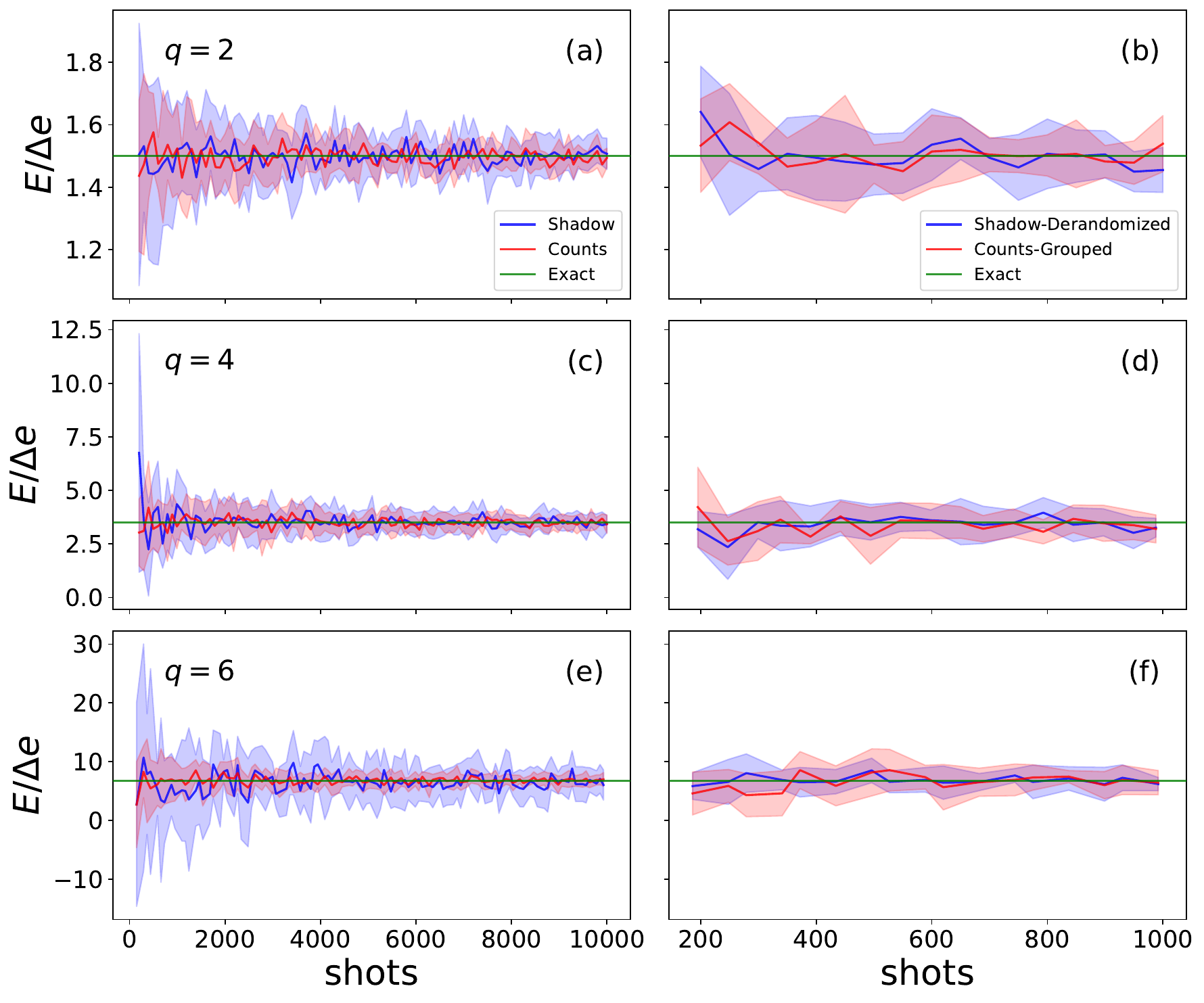}
    \caption{Comparison of the different methods in obtaining the energy associated with the projected pairing Hamiltonian on the even parity subspace for different numbers of qubits $q=2,4$ and $6$. (a), (c) and (e) correspond to the results using the straightforward version of the Classical Shadows method \cite{Hua20,Css23} together with a direct ``Counts'' method discussed in section \ref{sec:optimization}. In contrast, panels (b), (d), and (f) are associated with the optimized version of both methods: ``Shadow-Derandomized'' \cite{Hua21,Csd23} and Counts-grouped, which is the same ``Counts'' method but grouping qubit-wise commutative observables \cite{Dcg23}. In all cases, the shaded areas correspond to the error bars on the results obtained by performing 10 times the same calculations with a given number of shots. Note that here $E$ denotes only the numerator of the expression in Eq. \eqref{eq:obssr}.}
    \label{fig:comparison_shadow_counting_shaded_area}
\end{figure*}

Illustrations of the expectation values of the energies obtained after projection on the even parity Hilbert space are given respectively for $q=2$ (a), $4$ (c), and $6$ (e) using a straightforward application of the Classical Shadows method (blue-shaded area), i.e., without derandomization. We observe that the energy properly converges to the expected value, which validates the present protocol of classical shadows with symmetry restoration. We further illustrate this protocol below for particle number and total spin symmetry restoration.  

\subsection{Particle number projection}
\label{sec:projection_pn_shadow}

We now consider the projector associated with the particle number case. For this, we consider implicitly a JWT transformation between particles and qubits. Then, the particle number operator $N$ identifies with the operator defined in Eq. (\ref{eq:partnumb}), which is diagonal in the computation basis and whose eigenvalues identify with the number of ones in a basis state. As discussed in \cite{Rui22,Lac23,Rui23}, the associated projector for a given particle number $n_0$ can be written as: 
\begin{equation}
    {\cal P}_{n_0} = \frac{1}{q+1}\sum_{k=0}^{q}e^{2\pi ik\left(N-n_0\right)/\left(q+1\right)} = \sum_{k} g_{k}e^{i\phi_{k}N},
    \label{eq:projector_number_particles}    
\end{equation}
with $g_{k}=\frac{1}{q+1}e^{-i\phi_{k}n_0}$, $\phi_{k}=\frac{2\pi k}{q+1}$ and $q$ the maximum possible particle number. In our case, $q$ equals the number of qubits. Using the decomposition in terms of phase gates $Q_j\left(\phi\right)$: 
\begin{eqnarray}
Q_j (\phi) &=& \left[ 
\begin{array}{cc}
   1  & 0 \\
   0  & e^{i\phi}
\end{array}\right]_j, 
\end{eqnarray}
and using $e^{i\phi N} = \bigotimes_{j=0}^{q-1} Q_j\left(\phi\right)$, we deduce: 
\begin{equation}
    {\cal P}_{n_0} = \sum_{k}g_{k} \bigotimes_{j=0}^{q-1} Q_j\left(\phi_k\right). 
\end{equation}
Following the procedure in section \ref{sec:general_discussion_method}, we directly apply this operator to each snapshot in a post-processing step. The projected expectation value of an observable $O$ from Eq. \eqref{eq:expectation_observable} over a single snapshot $\hat{\rho}^{(n)}$ becomes:
\begin{equation}
    \Tr\left(O{\cal P}_N\hat{\rho}^{(n)}\right) = \sum_k g_k \prod_{j=0}^{q-1}\Tr\left[ P_j Q_j\left(\phi_k \right)\left(3r^{(n)}_j-I_j\right)\right].
    \label{eq:trace_projection_pn_first_form}
\end{equation}
Expressing the phase gates in terms of Pauli matrices, i.e.:
\begin{equation}
    Q_j\left(\phi\right)=e^{\lambda/2}\left[\cos\left(\phi/2\right)I_j  -iZ_j \sin\left(\phi/2\right)\right],
\end{equation}
we get: 
\begin{multline}
    \Tr\left(O {\cal P}_N\hat{\rho}^{(n)}\right) = \sum_{k=0}^{n} g_k e^{in\phi_k/2}\\
    \prod_{j=0}^{q-1}
    \left\{\cos\left(\phi_k/2\right)\Tr\left[P_j\left(3r^{(n)}_j-I_j\right)\right]\right.\\
    \left.-i\sin\left(\phi_k/2\right)\Tr\left[P_jZ_j\left(3r^{(n)}_j-I_j\right)\right]\right\}.
    \label{eq:trace_projection_particle_number}
\end{multline}

\begin{figure}[htbp]
    \centering
    \includegraphics[scale=0.55]{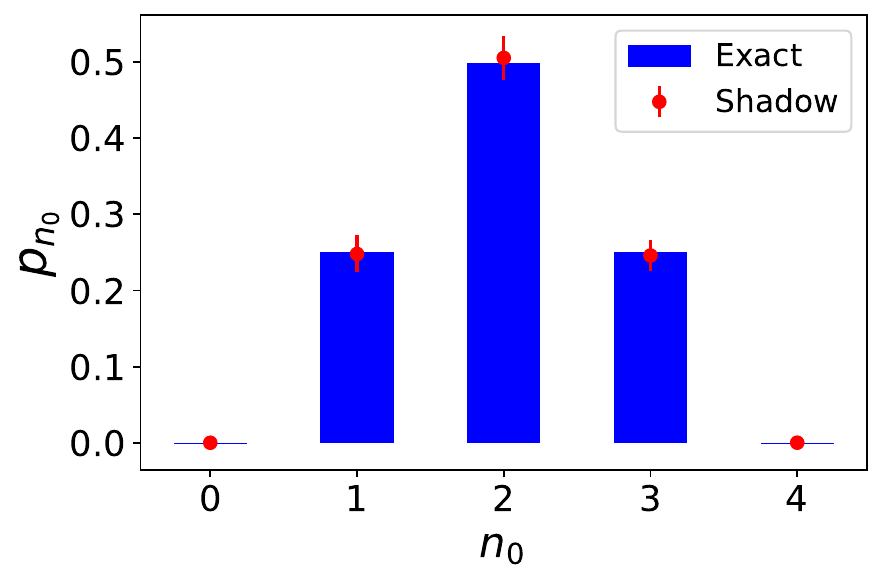}
    \caption{Illustration of the probability $p_{n_0}$ obtained by combining projection on particle number with the classical shadow procedure delineated in section \ref{sec:projection_pn_shadow} for $n_0=0,\cdots,4$. We consider here a Gaussian state encompassing the register $k=0,\dots,2^{q}-1$, where $q=4$. This state takes the form $|\psi_G\left(k\right) \rangle = \frac{1}{\mathcal{N}} \sum_k e^{-\frac{1}{2}  \left(\frac{k-\mu}{\sigma}\right)} |k\rangle$, with a median $\mu = \left(2^{q}-1\right)/2$ and a standard deviation $\sigma = \mu/3$. $\mathcal{N}$ is a normalization constant. The results obtained from the classical shadows were acquired using $10^4$ shots, repeated 50 times. The red data points and their corresponding error bars denote the mean value and standard deviation derived from these 50 trials.}
    \label{fig:particle_projection_shadow}
\end{figure}
Therefore, we see in this example that projected properties can be obtained using the classical shadow at the price of enlarging the set of Pauli strings schematically:
\begin{eqnarray}
    \{ P_j \} &\longrightarrow& \{ O_j \} \equiv \{P_j, ~ P_jZ_j \}. \label{eq:enlargeP}
\end{eqnarray}
This is also the enlarging used in the parity case. Note that the resulting number of Pauli strings can be much lower than twice the original number $M$ of $P_j$ due to the properties of Pauli matrices. 

 As mentioned before, having an estimate for the set of values $\Tr\left[ P_j R_j\left(\phi_k\right)\left(3r^{(n)}_j-I_j\right)\right]$, as depicted in Eq. (\ref{eq:trace_projection_pn_first_form}), provides us with the means to get projections onto all possible particle numbers $n_0=0,q$ using the same set of snapshots. 
 This is feasible as the information about $n_0$ is purely encoded in the coefficients $g_k$. Hence, it is only necessary to compute all products once to achieve the projected expectation value of the observable $O$ for any given particle number $n_0$. As a demonstration, Fig. \ref{fig:particle_projection_shadow} presents the probabilities of the projection over different particle numbers $n_0$ for a Gaussian state prepared on a quantum register with $q=4$ qubits. 

\begin{figure}[htbp]
    \centering
    \includegraphics[scale=0.55]{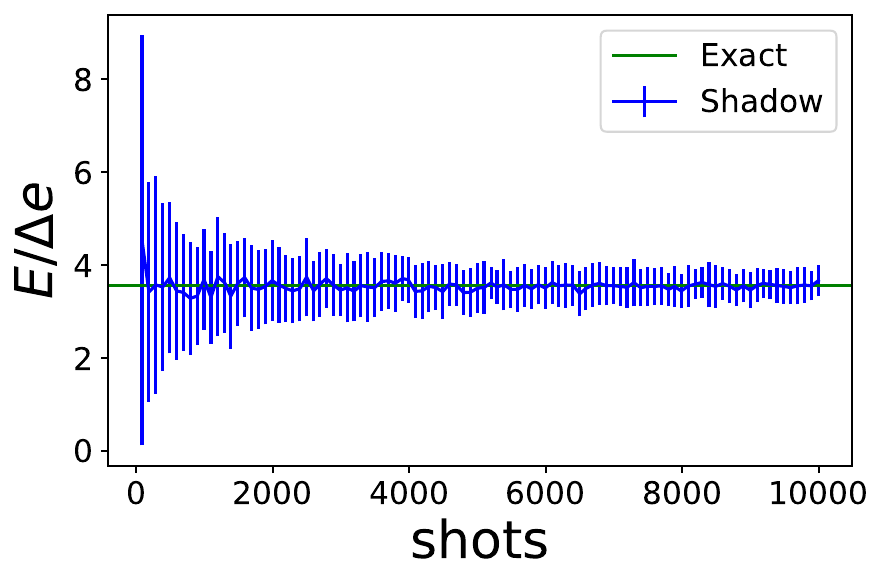}
    \caption{Convergence of the energy obtained after projection on particle number, employing the same symmetry breaking wave function as in Fig. \ref{fig:particle_projection_shadow}, with the pairing Hamiltonian. The initial state is prepared on $q=4$ qubits and projected 
    on $n_0=2$. Note that we directly encode pairs of particles on qubits; thus, $n_0$ identifies with the number of pairs. The results were obtained from the Classical Shadows method by repeating the process with the same number of \textit{shots} $50$ times. The blue data points and their corresponding error bars represent the mean outcome and the standard deviation derived from these $50$ iterations, respectively. Despite significant fluctuations, we can observe a convergence of the average energy value to the exact projected energy, even with small numbers of shots. Note that here $E$ denotes again only the numerator of the expression in Eq. \eqref{eq:obssr}.}
    \label{fig:projected_energy}
\end{figure}

Finally, Fig. \ref{fig:projected_energy} illustrates the projected energy of the wave function employed in Fig. \ref{fig:particle_projection_shadow}, in conjunction with the pairing Hamiltonian within the subspace with $N=2$ pairs. This figure was obtained using the methodology in \ref{sec:general_discussion_method} and shows that the correct energy is obtained as the number of shots increases. 

\subsection{Total Spin projection with classical shadow}
\label{sec:projection_spin_shadow}

\begin{figure}[htbp]
    \centering
    \includegraphics[scale=0.55]{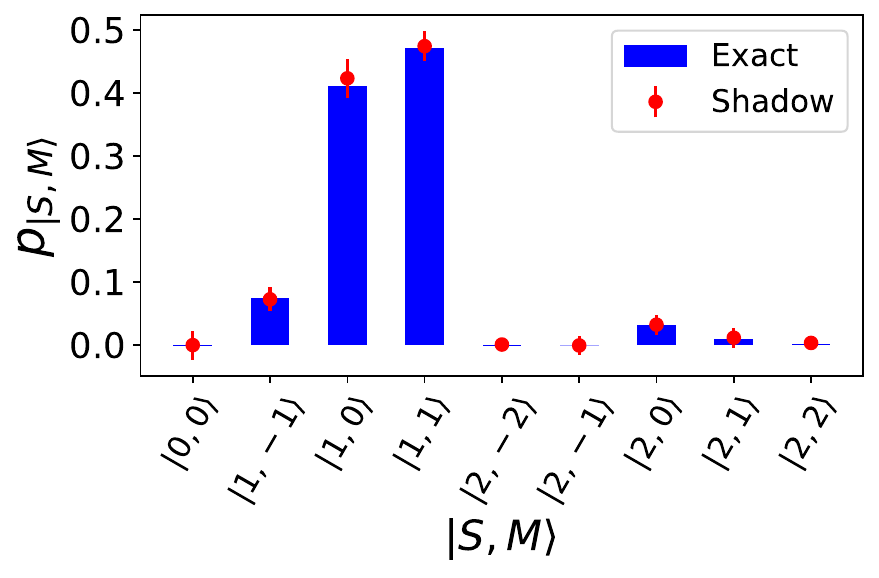}
    \caption{Illustration of the projection onto different basis states, denoted as $|s,m\rangle$, for a system composed of 4 qubits. The red dots are obtained using the classical shadow procedure, as detailed in Section \ref{sec:projection_spin_shadow}. We used a Gaussian state situated within the eigenspace of the $S^2$ operator. To be specific, given the $S^2$ operator's decomposition as $S^2 = UDU^\dagger$, we use $|\psi\rangle = U |\psi_G\left(k\right)\rangle$, where $|\psi_G\left(k\right)\rangle$ is the state used in Fig. \ref{fig:particle_projection_shadow}. The results derived from the Classical Shadows technique were acquired using $10^4$ shots, repeated 50 times. The red data points and their respective error bars symbolize the mean and standard deviation calculated from these 50 trials, respectively. Results displayed here are obtained with $n_p=10$ points to discretize each integral, and the small difference seen between the histogram and the red dots are consistent with the error induced by the discretization of the integral ($\sim 1$\% for $n_p=10$) .}
    \label{fig:spin_projection_shadow}
\end{figure}
We consider here the case of projection onto total spin symmetry that introduces greater complexity because we consider a symmetry operator ${\bf S}^2$ that contains two-qubit operations and that the computational basis is not an eigenbasis of this operator. Here, ${\bf S} = (S_x, S_y, S_z)$ denotes the total spin operator. A possible starting point is to utilize the ${\bf S}^2$ operator in a projector analogous to Eq. \eqref{eq:projector_number_particles} as was done in Refs. \cite{Siw21,Lac23}. Such expression of the projector prevents the operator $e^{i\phi_k {\bf S}^2}$ from being decomposed into a tensor product of single-qubit operators, thereby hindering our application of the orthogonality of the Pauli operators to streamline computations. 

To circumvent this issue, we adopt the integral form of the projector for the spin states $|s,m\rangle$ \cite{Rin80,She19}, which possesses total spin $s\left(s+1\right)$ and projection on the z-axis $m$ (where $\hbar=1$): 
\begin{widetext}
\begin{eqnarray}
{\cal P}_{s,m} &=& | s, m \rangle \langle s, m | = 
\frac{2s+1}{8 \pi^2} \int_0^{2\pi}  \int_0^\pi   \int_0^{2\pi}  
        \sin(\beta) \left(D^{s}_{m,m}(\alpha,\beta, \gamma) \right)^* 
        e^{-i \alpha S_z} e^{-i \beta S_y} e^{-i \gamma S_z}d\alpha d\beta d\gamma,
        \label{eq:total_spin_projector}
\end{eqnarray}
where $D^{s}_{m,m}(\alpha,\beta, \gamma)$ is the $(s,m)$ component of the Wigner D--matrix $D(\alpha,\beta, \gamma)$. 
This form was already used in the quantum computing context in \cite{Tsu20,Tsu22,Sek22}. Discretizing the integrals and using the relations:
\begin{equation}
    S_y = \frac{1}{2}\sum_{j=0}^{q-1} Y_j, \qquad S_z = \frac{1}{2}\sum_{j=0}^{q-1} Z_j,
\end{equation}
we obtain the following form of the spin projector:
\begin{eqnarray}
    {\cal P}_{s,m} &=&  \Delta \Omega(s) \sum_{k,l,p} \sin(\beta_l) \left[D^{s}_{m,m}(\alpha_k,\beta_l, \gamma_p)\right]^* 
    \bigotimes_{j=0}^{q-1} R^j_z\left(\alpha_k\right) R^j_y\left(\beta_l\right) R^j_z\left(\gamma_p\right) ,
    \label{eq:spin_projector_discrete}
\end{eqnarray}
where, $\Delta \alpha$, $\Delta \beta$, and $\Delta \gamma$ denote the step sizes for the respective angles $\alpha$, $\beta$, and $\gamma$, i.e., $\alpha_k = k\Delta \alpha$, $\beta_l = l\Delta \beta$, and $\gamma_p = p\Delta \gamma$ with $k,l,p=0,\dots,n_p$, and $n_p$ indicates the number of points considered for each integral. Here $\Delta \Omega(s)$ is a factor that includes the different mesh steps stemming from the integral discretization:
\begin{eqnarray}
    \Delta \Omega(s) &=& \frac{2s+1}{8 \pi^2}\Delta \alpha \Delta \beta \Delta \gamma. \label{eq:intfactor}
\end{eqnarray}  
The number of terms in the $k,l,p$ sum -- i.e., the number of $ R^j _z\left(\alpha_k\right) R^j_y\left(\beta_l\right) R^j_z\left(\gamma_p\right)$ operators factored into the calculation -- is directly proportional to the precision of the discretization. 

The projected expectation value of an observable $O = \bigotimes_{j=0}^{q-1} P_j$ over a single snapshot $\hat{\rho}^{(n)}$ becomes: 
\begin{eqnarray}
    \Tr\left[ O {\cal P}_{s,m}\hat{\rho}^{(n)}\right] &=& \Delta \Omega(s)  
    \sum_{k,l,p} \sin(\beta_l) \left(D^{s}_{m,m}(\alpha_k,\beta_l, \gamma_p)\right)^* \prod_{j=0}^{q-1} {\rm Tr}\left[P_j B_j\left(\alpha_k,\beta_l, \gamma_p\right) \left( 3r^{(n)}_j - I_j \right) \right], 
    \label{eq:trace_projection_spin}
\end{eqnarray}
where we have defined the operator $B_j(\alpha_k,\beta_l, \gamma_p)$ acting on the qubit $j$ as:
\begin{eqnarray}
    B_j(\alpha_k,\beta_l, \gamma_p) &\equiv&  R^j _z\left(\alpha_k\right) R^j_y\left(\beta_l\right) R^j_z\left(\gamma_p\right)  = \sum_{Q_j=I_j,X_j,Y_j,Z_j} c_{Q}(\alpha_k,\beta_l, \gamma_p) Q_j,
\end{eqnarray}
with the set of coefficients $c_Q$ given by:
\begin{eqnarray}
\left\{ 
\begin{array}{l}
{\displaystyle c_I (\alpha,\beta, \gamma) =} 
{\displaystyle \cos\left(\alpha\right)\cos\left(\beta\right)\cos\left(\gamma\right)}
{\displaystyle - \frac{1}{4}\sin\left(\alpha\right)\cos\left(\beta\right)\cos\left(\gamma\right)}
\\
{\displaystyle c_X (\alpha,\beta, \gamma) =} 
{\displaystyle \frac{i}{4}\left[\sin\left(\alpha\right)\sin\left(\beta\right)\cos\left(\gamma\right)\right.}
{\displaystyle - \left.\cos\left(\alpha\right)\sin\left(\beta\right)\sin\left(\gamma\right)\right]} \\
{\displaystyle c_Y (\alpha,\beta, \gamma) =} 
{\displaystyle -\frac{i}{2}\left[\vphantom{\frac{1}{4}}\cos\left(\alpha\right)\sin\left(\beta\right)\cos\left(\gamma\right)\right.}
{\displaystyle + \left.\frac{1}{4}\sin\left(\alpha\right)\sin\left(\beta\right)\sin\left(\gamma\right)\right]}
\\
{\displaystyle c_Z (\alpha,\beta, \gamma) =} 
{\displaystyle -\frac{i}{2}\left[\cos\left(\alpha\right)\cos\left(\beta\right)\cos\left(\gamma\right)\right]}
{\displaystyle + \left.\sin\left(\alpha\right)\cos\left(\beta\right)\cos\left(\gamma\right)\right]}
\end{array}\right. \qquad .
\end{eqnarray}
\end{widetext}
The above expressions demonstrate again that the projection can be incorporated using the classical shadow by enlarging the set of Pauli strings to be evaluated. A consideration, similar to the one presented in the particle number projection case, can be made for the total spin projection: Computing the desired expectation of all Pauli strings once provides access to the projected expectation values for all combinations of the $(s,m)$ parameters. We present in Fig. \ref{fig:spin_projection_shadow} the amplitude decomposition obtained using the Classical Shadows technique for a specific quantum state, using the $\left({\bf S}^2, S_z\right)$ eigenbasis denoted by $|s,m\rangle$. Note that projection on $S_z$ that is performed inside the projection on ${\bf S}^2$ is in practice equivalent to the projection on particle number discussed in the previous section (see also Eq. (\ref{eq:partnumb}) and Ref. \cite{Lac23}). As we see again from Fig. \ref{fig:spin_projection_shadow}, the procedure discussed here using Classical Shadows successfully reproduces the correct amplitudes for any $(s,m)$ values. 

\section{Analysis of the classical shadows performance and comparison with other post-processing techniques}
\label{sec:optimization}

One of the attractive aspects of the Classical Shadows technique is its performance in terms of the number of measurements versus precision on a given set of observables.  Specifically, it has been proven that, using a classical shadow of $M$ events, it is possible to predict $L$ arbitrary linear functions $\Tr\left(O_1\rho\right), \dots, \Tr\left(O_L\rho\right)$ up to an additive error $\epsilon$ if:
\begin{equation}
    M \geq \mathcal{O} \left(\left(\log L\right) \left({\rm max} \left\{\left|\left|O_i\right|\right|^2_{{\rm shadow}}\right\}_{i=1,\dots,L} \right)/ \epsilon^2 \right).
\end{equation} 
The shadow norm $\left|\left|O_i\right|\right|^2_{{\rm shadow}}$ depends on the unitary ensemble that is chosen \cite{Hua20}.

We already have shown, for instance, in Figs. \ref{fig:parityprob_shadow} and \ref{fig:comparison_shadow_counting_shaded_area}, practical illustrations of the convergence properties of the shadow estimates as a function of the number of shots. The present section aims to investigate this aspect further and see (i) if some optimization of the shadow can be further investigated to reduce the required number of shots given a certain precision and (ii) if the classical shadow method is competitive with respect to alternative techniques able to perform symmetry restoration using classical computer post-processing.

We can compare the technique laid out in this paper with a ``direct counts'' one by computing the expectation values of the operators $O'_\alpha$ in $H\mathcal{P}=\sum_\alpha c_\alpha O'_\alpha$ with the latter method. All the results presented (for the Classical Shadows and Counts techniques) in Fig. \ref{fig:comparison_shadow_counting_shaded_area} were obtained by directly computing the expectation values of the observables $\langle O'_\alpha \rangle$ instead of using the product of the traces as given in Eq. (\ref{eq:expectation_observable}). We expect the non-optimized version of this method to be equivalent to the direct use of the trace formula. Given a symmetry-breaking state $| \Psi \rangle$ prepared on the qubit register, we perform a change of basis from the original computational basis to a new basis where $O'_\alpha$ is diagonal prior to the measurement (see also discussion in \cite{Ayr23}, Table 3, where the different unitary transformations to perform are discussed). 

Provided that the set of values $\{ \langle O'_\alpha \rangle \}$ are retrieved from the different set of measurements. The expectation value of the symmetry restored state can be obtained simply by making the proper linear combination of them on a classical computer. Note that the norm of the projected state is obtained by considering the identity as the observable. This technique is standardly used in the VQE algorithm and was used already for symmetry restoration, for instance, in \cite{Kha21,Tsu20}. We call this method ``Direct counts'' or simply the ``Counts'' method. 

We focus below on comparing the Classical Shadows application and the direct counts method for the parity projection case. In panels (a), (c), and (e) of Fig. \ref{fig:comparison_shadow_counting_shaded_area}, results obtained using the Direct counts are reported for the same number of shots as in the Classical Shadows case, including the deviation with respect to the mean values. We observe in this figure that the alternative method clearly outperforms the classical shadow in terms of precision, especially as the number of qubits increases.  

The results reported up to now correspond to a straightforward application of both the Classical Shadows and direct counts techniques. In both cases, the possibility of optimizing the approaches and reducing the number of shots to achieve a given precision has been proposed. In the present work, we have explored some of these optimizations: 

\begin{itemize}
    \item {\bf Optimization of the Classical Shadows:} We found that the derandomization technique of Ref. \cite{Css23} leads to a significant reduction of the number of shots.
    \item {\bf Optimization of the direct counts:} For this approach, we used the grouping technique of Ref. \cite{Dcg23}. The grouping algorithm used is based only on the Qubit-wise commutation of different observables. Other forms of commutation can be taken into account to optimize the number of measurements further using grouping techniques \cite{Yem20,Izm19,Hug21} (see also \cite{Dcg23}). To determine the groups in the grouping method, the results presented here are obtained using the grouping coloring heuristic ``Recursive Largest First'' (rlf), which is a variant of the largest first coloring heuristic \cite{Dcg23}.
\end{itemize}

We show in panels (b), (d), and (f) of Fig. \ref{fig:comparison_shadow_counting_shaded_area} a comparison of the projected energy obtained using both methods of optimization. In all cases, we observe an improved convergence, together with a much smaller deviation of the results to the exact energy. We also 
see that both techniques become comparable, whatever the number of qubits, showing that the classical shadow can indeed be competitive once using the derandomization technique.


\section{Conclusion}
In our pursuit to discover alternative methods for minimizing the resources required for the symmetry restoration process, we've delved into the Classical Shadows paradigm. We show here that projection onto certain symmetries, namely parity, particle number, or total spins, can be made by performing classical post-processing using the classical shadow approach. This become possible by representing symmetry projection operators as weighted sums of tensor products of single qubit operators--we showed how the orthogonality of the Pauli operators under the Hilbert-Schmidt inner product facilitates efficient computation of projected expectation values. 

We finally analyze critically how the precision of the method can be improved using 
the derandomization approach. The accuracy of predicting observables for symmetry-restored states 
is compared to alternative post-processing methods, also applied with optimization. We show that the 
classical shadow approach can effectively compete with other methods while having more potential.

\subsection{Acknowledgments }
This project has received financial support from the CNRS through
the 80Prime program  and the AIQI-IN2P3 project. This work is part of 
HQI initiative (www.hqi.fr) and is supported by France 2030 under the French 
National Research Agency award number "ANR-22-PNQC-0002".
We acknowledge
the use of IBM Q cloud as well as use of the Qiskit software package
\cite{Qis21} for performing the quantum simulations.

\bibliographystyle{unsrt}

\end{document}